\documentstyle[prl,aps,epsf]{revtex}
\begin{document}
\draft
\tighten
\title{\hfill {\small DOE-ER-40757-093} \\
\hfill {\small UTEXAS-HEP-97-2} \\
\hfill  {\small MSUHEP-70205} \\
\hfill \\ 
Photon--neutrino interactions} 

\author{Duane A. Dicus} 
\address{Center for Particle Physics and Department of Physics \\
 University of Texas, Austin, Texas 78712}

\author{Wayne W. Repko} 
\address{Department of Physics and Astronomy \\  
Michigan State University, East Lansing, Michigan 48824}
\date{\today}
\maketitle

\begin{abstract}
The cross sections for the processes $\gamma \nu\rightarrow \gamma
\gamma \nu$, $\gamma\gamma\rightarrow\gamma\nu\bar{\nu}$ and
$\nu\bar{\nu}\rightarrow\gamma\gamma\gamma$ are calculated with
the aid of an effective Lagrangian derived from the Standard model. 
These cross sections are shown to be much larger
than the elastic cross section $\sigma(\gamma\nu\rightarrow\gamma\nu)$ for
photon energies $\omega > 1$\,keV. Possible astrophysical implications are
discussed.
\end{abstract}

\section{Elastic processes}

Photon-neutrino scattering is of potential interest in astrophysical processes.
The various $\gamma\nu$ reaction cross sections, which are expected to be small 
due to the
weak interaction, are further suppressed because of the vector--axial vector
nature of the weak coupling \cite{mjl,cy,liu}. For the channel $\nu\bar{\nu}
\rightarrow\gamma\gamma$, Yang's theorem \cite{yang}
implies a vanishing decay amplitude to order $G_F$
when the neutrinos are massless. This means that the amplitude in any channel
contains
additional factors of $\omega /m_W$, where $\omega$ is the photon energy and
$m_W$ is the $W$ mass. If the neutrinos have a mass $m_{\nu}$, the amplitude is 
reduced by factors of $m_{\nu}/m_W$ \cite{gell}. 

In the case of massless neutrinos, the amplitude for $\gamma\nu\rightarrow 
\gamma\nu$ in the Standard model is \cite{dr93}, to leading order in 
$m_W^{-2}$,  
\begin{equation}
{\cal M}_{\lambda\lambda^{\prime}} = \frac{1}{2\pi}\frac{g^2\,\alpha}{m_W^4}
\left[1 + \frac{4}{3}\ln\left(\frac{m_W^2}{m_e^2}\right)\right]
\cos(\theta/2)\left(\,s^2(1 + \lambda\lambda^{\prime}) + 
\frac{st}{2}(1 + \lambda)(1 + \lambda^{\prime})\right)\,,
\end{equation}
where $\lambda$ and $\lambda^{\prime}$ are the photon helicities, $s = 4
\omega^2$, $t = -2\omega^2(1 - \cos\theta)$ and $\theta$ is the scattering 
angle. 
Notice that there are no inverse powers of $m_e$ in this expression and hence
the scale is set by $m_W$. As a consequence, all $2\rightarrow 2$ $\gamma \nu$
channels can be calculated using the effective Lagrangian
\begin{equation}
{\cal L}_{\rm eff} = \;\frac{1}{4\pi}\frac{g^2\alpha}{m_W^4}
\left[1 + \frac{4}{3}\ln\left(\frac{m_W^2}{m_e^2}\right)\right] 
\left[\bar{\psi}\gamma_{\nu}(1 + \gamma_5)
(\partial_{\mu}\psi) - (\partial_{\mu}\bar{\psi})\gamma_{\nu}(1 +
\gamma_5)\psi\right]F_{\mu\lambda}F_{\nu\lambda}\;,
\end{equation}
where $F_{\mu\nu}$ is the photon field tensor and $\psi$ is the neutrino field.
The resulting cross section for elastic scattering is
\begin{equation} \label{elas}
\sigma(\gamma\nu\rightarrow\gamma\nu) = \frac{3}{4}\frac{G_F^{\,2}\,\alpha^2}
{\pi^3}\left[1 + \frac{4}{3}\ln\left(\frac{m_W^2}{m_e^2}
\right)\right]^2\left(\frac{\omega}{m_W}\right)^4\omega^2\,.
\end{equation}
Despite the enhancement of the $\ln^2\left(m_W^2/m_e^2\right)$ factor,
this cross section is exceedingly small and likely to be of little practical
importance in astrophysics.

\section{Inelastic processes}

The source of the large suppression in the $\gamma\nu$ elastic amplitude is the
Yang theorem prohibition of a two photon coupling to a $J = 1$
state. There is no similar restriction on the coupling of three photons. This
suggests an examination of the inelastic process  $\gamma\nu\rightarrow
\gamma\gamma\nu$ to determine if the scale of the loop integrals resulting
from the $W$ and $Z$ exchange diagrams represented by Fig.\,1 is set by
the electron mass rather than $m_W$. This turns out to be the case, and
one can obtain an effective Lagrangian of the form
\begin{equation} \label{l3gam}
{\cal L}_{\rm eff} = 4\frac{G_F\,a}{\sqrt{\displaystyle 2}}
\frac{\alpha^{3/2}}{\sqrt{\displaystyle 4\pi}}
\frac{1}{m_e^4}\left[\frac{5}{180}\left(N_{\mu\nu}F_{\mu\nu}
\right)\left(F_{\lambda\rho}F_{\lambda\rho}\right) - \frac{14}{180}N_{\mu\nu}
F_{\nu\lambda}F_{\lambda\rho}F_{\rho\mu}\right]\,,
\end{equation}
where $N_{\mu\nu}$ is
\begin{equation} 
N_{\mu\nu} = \partial_{\mu}\left(\bar{\psi}\gamma_{\nu}(1 + \gamma_5)\psi
\right) 
- \partial_{\nu}\left(\bar{\psi}\gamma_{\mu}(1 + \gamma_5)\psi\right)\,,
\end{equation}
and $a = 1 - \case{1}{2}(1 - 4\sin^2\theta_W)$. 
The numerical factors $5/180$ and $-14/180$ are familiar from the
Euler-Heisenberg \cite{e-h} expansion of the photon--photon scattering
amplitude. This occurs because after Fierz rearranging the $W$ contribution and
replacing the gauge boson propagators by their masses,
the amplitude in any channel for either $W$ or $Z$ 
exchange reduces to a four photon amplitude with one photon polarization 
vector replaced by the neutrino current.
The $Z$ contribution, given by the second factor in $a$, is
numerically small $\approx .04$.  Using Eq.\,(\ref{l3gam}), the double
differential cross section $d^2\sigma/d\omega_1 d\omega_2$ for the energy
distribution of the final photons in the neutrino-photon center of mass is
\begin{eqnarray} \label{dsigdw1dw2}
\frac{d^2\sigma}{d\omega_1 d\omega_2} & = & \frac{G_F^2\,a^2\alpha^3}
{\pi^4 m_e^8}
\frac{\omega^3}{30,375}\Bigl[-\frac{1}{2}\left(3614\,\omega_1^2\omega_2^2
(\omega_1 + \omega_2) + 2085\,\omega_1\omega_2(\omega_1^3 + \omega_2^3)\right.
\nonumber \\ 
&   &\left.+ 695\,(\omega_1^5 + \omega_2^5)\right) + 
\omega\left(5372\,\omega_1\omega_2(\omega_1^2 + \omega_2^2) + 
8103\,\omega_1^2\omega_2^2 + 2085\,(\omega_1^4 + \omega_2^4)\right) \nonumber 
\\ 
&   &- \omega^2\left(15,552\,\omega_1\omega_2(\omega_1 + \omega_2) 
+ 6067\,(\omega_1^3 + \omega_2^3)\right) + \omega^3\left(\mbox{\rule
{0pt}{11pt}}21,750\,\omega_1\omega_2\right. \nonumber \\ 
&   &\left.+ 11,063\,(\omega_1^2 + \omega_2^2)\right)
- \frac{21,055}{2}\,\omega^4(\omega_1 + \omega_2) + 3794\,\omega^5\,
\Bigr]\,. 
\end{eqnarray}
The cross section for $\gamma\nu\rightarrow \gamma\gamma\nu$ can be obtained by
integrating Eq.\,(\ref{dsigdw1dw2}) over the region $(\omega - \omega_1) \leq
\omega_2 \leq \omega$\,, $0 \leq \omega_1 \leq \omega$ and is
\begin{equation} \label{inelas}
\sigma(\gamma\nu\rightarrow\gamma\gamma\nu) = \frac{262}{127,575}
\frac{G_F^{\,2}\,a^2\,\alpha^3}{\pi^4}\left(\frac{\omega}{m_e}\right)^8
\omega^2\,.
\end{equation}

Further details of the scattering process are presented
in Fig.\,2. By retaining the polarization vector of one of the final state
photons, it is possible to obtain the cross section for this photon to be
produced with either positive or negative helicity. This is illustrated in the 
left panel of Fig.\,2 by the dashed and dot-dashed lines. The solid line in this
panel is the angular distribution for unpolarized scattering and $\theta$ is the
angle between the outgoing photon and the incident photon. The difference 
between the positive and negative helicity cross sections, which are fifth order
polynomials in $\cos\theta$, results in the polarization $P(\theta)$ illustrated
in the right panel of Fig.\,2. The existence of a net circular polarization is
possible because the weak interaction violates parity.

It is also of interest to examine the annihilation channel $\gamma\gamma
\rightarrow \gamma\nu\bar{\nu}$, which provides an energy loss
mechanism for stellar processes \cite{cm,vhs,km}. In this case, 
the double differential cross
section $d^{\,2}\sigma/\sin\theta d\theta d\omega^{\prime}$ is
\begin{equation}\label{dbl}
\frac{d^{\,2}\sigma}{\sin\theta d\theta d\omega^{\prime}}  =  
\frac{G_F^{\,2}\,a^2\,\alpha^{\,3}}{\pi^4m_e^8}\,
\frac{\omega^3\omega^{\prime\;3}(\omega - \omega^{\prime})}{48,600}
\biggl[2224\omega^2 - \omega(592\omega + 520\omega^{\prime})\sin^2\!\theta 
+ 139\omega^{\prime\;2}\sin^4\!\theta\biggr]\,,
\end{equation}
where $\theta$ is the scattering angle of the final photon in the center of
mass, $\omega^{\prime}$ is its energy and $\omega$ is the initial
photon energy. When integrated, Eq.\,(\ref{dbl}) gives the total cross section
\begin{equation} \label{ggnng}
\sigma(\gamma\gamma\rightarrow\gamma\nu\bar{\nu}) = \frac{2,144}{637,875}
\frac{G_F^2\,a^2\,\alpha^3}{\pi^4}\left(\frac{\omega}{m_e}\right)^8\,\omega^2\,.
\end{equation}
Unlike $\gamma\nu\rightarrow \gamma\gamma\nu$, the final
photon in the annihilation channel does not acquire any circular polarization.

The neutrino annihilation channel $\nu\bar{\nu}\rightarrow\gamma\gamma\gamma$ is
potentially important in supernova calculations. Here, the energy distribution
of two of the final photons in the neutrino center of mass is
\begin{eqnarray}
\frac{d^2\sigma}{d\omega_1 d\omega_2} & = & \frac{G_F^2\,a^2\alpha^3}
{\pi^4 m_e^8}
\frac{8\omega^4}{18,225}\Bigl[139\left((\omega_1^4 + \omega_2^4) +
3\omega_1^2\omega_2^2 + 2\omega_1\omega_2(\omega_1^2 + \omega_2^2)\right)
\nonumber \\
&   &- 4\omega\left(139\,(\omega_1^3 + \omega_2^3)
 + 315\,\omega_1\omega_2(\omega_1 + \omega_2)\right) + 2\omega^2
\left(491(\omega_1^2 + \omega_2^2) + 917\omega_1\omega_2\right) \nonumber \\ 
&   &- 852\,\omega^3(\omega_1 + \omega_2) + 287\omega^4\Bigr]\,,
\end{eqnarray}
which may be integrated to give
\begin{equation} \label{nnggg}
\sigma(\nu\bar{\nu}\rightarrow\gamma\gamma\gamma) = \frac{136}{91,125}
\frac{G_F^2\,a^2\,\alpha^3}{\pi^4}\left(\frac{\omega}{m_e}\right)^8\,\omega^2\,.
\end{equation}

\section{Discussion and conclusions}

A direct comparison of the elastic and inelastic cross sections, Eqs.\,
(\ref{elas}), (\ref{inelas}), (\ref{ggnng}) and (\ref{nnggg}), is given in 
Table 1. The $\omega^{10}$ behavior
of the inelastic cross sections versus the $\omega^6$ behavior of the elastic
cross section is evident as is the 14 orders of magnitude difference at $\omega
= 2m_e$. Certainly, the effective Lagrangian Eq.\,(\ref{l3gam}) provides an 
adequate description of the inelastic processes for $\omega < m_e$. The
application of Eq.\,(\ref{dbl}) to stellar energy loss is therefore completely
justified. At some point beyond $\omega = m_e$, 
the cross section ceases to grow as the tenth power, begins a
transition to a `soft' behavior and eventually decreases. The precise range of 
applicability of the power law is somewhat subjective in the sense that 
numerical factors resulting from the loop integrals which define ${\cal L}_
{\rm eff}$ are often included in the definition of the scale factor. Including 
the factor of 180 appearing in Eq.\,(\ref{l3gam}), the effective scale is 
$\sim 4 m_e$.

As a rough indication of the importance of the $\gamma\nu\rightarrow
\gamma\gamma\nu$ process in cosmology, consider the mean number of 
collisions in an expansion time $t$. Assuming that there is an effective scale
greater than $m_e$, Eq.\,(\ref{inelas}) can be written
\begin{equation}
\sigma(\gamma\nu\rightarrow\gamma\gamma\nu) = 2.0\times 10^{-53}T_{10}^{10}
\,\mbox{\rm cm$^2$} \,,
\end{equation}
where $T_{10}$ is the photon energy in units of $10^{10}K$ and $\omega \sim T$.
The mean number of collisions in this time \cite{peeb}, 
$\sigma(\gamma\nu\rightarrow\gamma\gamma\nu)n_{\nu}ct$, 
where $n_{\nu}$ denotes the neutrino density, is
\begin{equation}
\sigma(\gamma\nu\rightarrow\gamma\gamma\nu)n_{\nu}ct = 1.92\times
10^{-11}T_{10}^{11}\,.
\end{equation}
For this to be of order 1, $T_{10}\sim 9.4$ or $T \sim 15\,m_e$. Strictly
speaking, this result is beyond the scale $\sim 4 m_e$, and it is probably
necessary to treat the transition region more carefully in order to determine
the decoupling temperature. If this temperature were shown to be low enough,
the process $\gamma\nu\rightarrow\gamma\gamma\nu$ might be of some importance in
cosmological considerations. 
There is the  possibility, albeit remote, that the detection of 
circular polarization, which only occurs in parity violating processes, could
provide evidence for relic neutrino interactions. 

\acknowledgements

We would like to thank V. Teplitz, M. Einhorn and V. Zelevinsky for helpful 
conversations. We are indebted to A. V. Kuznetsov and N. V. Mikheev for drawing
our attention to Ref.\,\cite{vhs} and informing us of their work.
This research was supported in part by the U.S. Department of Energy under
Contract No. DE-FG013-93ER40757 and in part by the National Science Foundation 
under Grant No. PHY-93-07980.

\begin{table}[h]
\caption{$\sigma(\omega)$ in cm$^2$ for $\omega$ in keV}
\begin{center}
\begin{tabular}{|c|c|c|c|c|} \hline
\mbox{\rule[-4pt]{0pt}{14pt}}
$\omega$ &$\sigma(\gamma\nu\rightarrow\gamma\gamma\nu)$ &
$\sigma(\gamma\gamma\rightarrow\gamma\nu\bar{\nu})$ &
$\sigma(\nu\bar{\nu}\rightarrow\gamma\gamma\gamma)$ &
$\sigma(\gamma\nu\rightarrow\gamma\nu)$ \\ \hline
\mbox{\rule{0pt}{11pt}}
1    &$8.66\times 10^{-83}$ &$1.42\times 10^{-82}$  &$6.29\times 10^{-83}$ &
$2.05\times 10^{-84}$ \\
100  &$8.66\times 10^{-63}$ &$1.42\times 10^{-62}$  &$6.29\times 10^{-63}$ &
$2.05\times 10^{-72}$ \\
511  &$1.05\times 10^{-55}$ &$1.72\times 10^{-55}$  &$7.63\times 10^{-56}$ &
$3.65\times 10^{-68}$ \\
1022 &$1.08\times 10^{-52}$ &$1.77\times 10^{-52}$  &$7.85\times 10^{-53}$ &
$2.34\times 10^{-66}$ \\
\hline
\end{tabular}
\end{center}
\end{table}

\begin{figure}[h]
\hspace{0.9in}
\epsfysize=2.1in \epsfbox{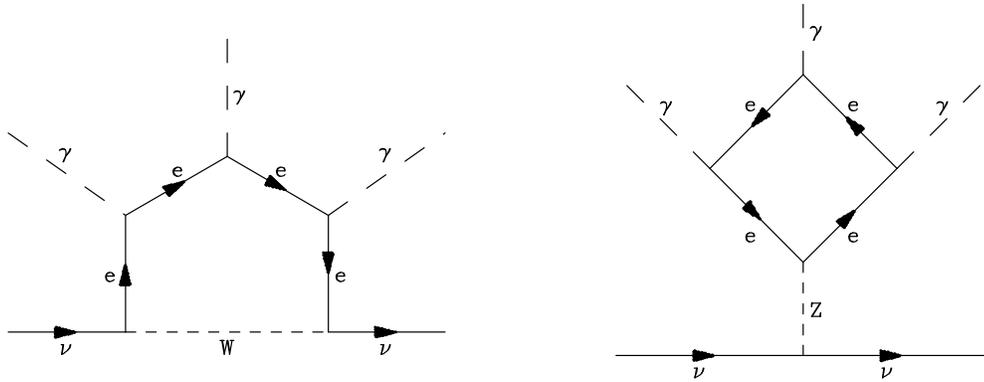}
\caption{Typical diagrams for the process $\gamma\nu\protect\rightarrow\gamma
\gamma\nu$ arising from $W$ (left) and $Z$ (right) exchange are 
shown. The complete set is obtained by permuting the photons.}
\end{figure}

\begin{figure}[h]
\hspace{1.5in}
\epsfysize=2.5in \epsfbox{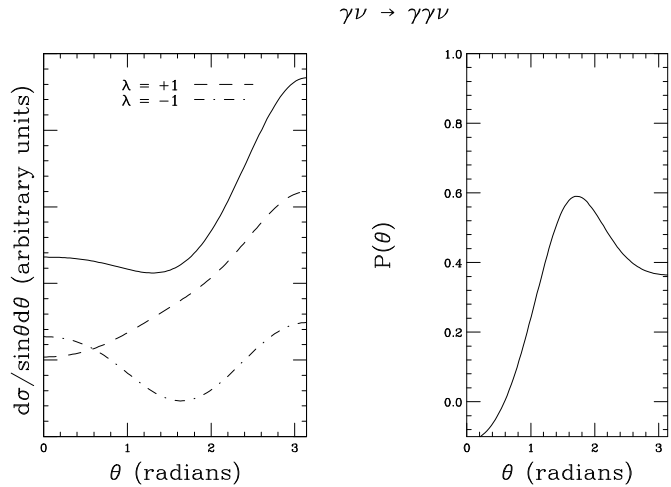}
\vspace*{.10in}
\caption{The differential cross section $d\sigma / \sin\theta d\theta$ and 
polarization
$P(\theta)$ are shown. In the left panel, the dashed line denotes the cross
section for the production of a positive helicity photon, the dot-dashed line
the cross section for the production of a negative helicity photon and the solid
line is  the sum of the two.}
\end{figure}


\begin{thebibliography}{99}
\bibitem{mjl} M. J. Levine, Nuovo Cimento {\bf 48A}, 67 (1967).
\bibitem{cy} V. K. Cung and M. Yoshimura, Nuovo Cimento {\bf 29A}, 557 (1975).
\bibitem{liu} J. Liu, Phys. Rev. D {\bf 44}, 2879 (1991).
\bibitem{yang} C. N. Yang, Phys. Rev. {\bf 77}, 242 (1950); L. D. Landau,
Sov. Phys. Doklady {\bf 60}, 207 (1948).
\bibitem{gell} M. Gell-Mann, Phys. Rev. Lett. {\bf 6}, 70 (1961).
\bibitem{dr93} D. A. Dicus and W. W. Repko, Phys. Rev. D {\bf 48}, 5106 (1993).
\bibitem{e-h} H. Euler, Ann. Phys. {\bf 26}, 398 (1936); W. Heisenberg and H.
Euler, Zeit. Phys. {\bf 98}, 714 (1936).
\bibitem{cm} H.-Y. Chiu and P. Morrison, Phys. Rev. Lett. {\bf 5}, 573 (1960).
\bibitem{vhs} N. Van Hieu and E. P. Shabalin, Sov. Phys. JETP {\bf 17}, 681
(1963). Our result, Eq. (\ref{dbl}), for the double differential cross section
differs from the one given in this paper.
\bibitem{km} A. V. Kuznetsov and N. V. Mikheev (unpublished).
\bibitem{peeb} P. J. E. Peebles, {\em Principles of Physical Cosmology},
Princeton University Press, Princeton, New Jersey (1993).
\end{thebibliography}
\end{document}